\documentclass[aps,prb,showpacs,twocolumn,superscriptaddress]{revtex4-1}
\usepackage{amsmath}
\usepackage{amssymb}
\usepackage{graphicx}
\usepackage{hyperref}
\usepackage{url}
\begin{document}
\title{Magnetic and Orbital Order in ($R$MnO$_3$)$_n$/($A$MnO$_3$)$_{2n}$ Superlattices Studied via a Double-Exchange Model with Strain}
\author{Shuai Dong}
\affiliation{Department of Physics, Southeast University, Nanjing 211189, China}
\affiliation{Department of Physics and Astronomy, University of Tennessee, Knoxville, Tennessee 37996, USA}
\affiliation{Materials Science and Technology Division, Oak Ridge National Laboratory, Oak Ridge, Tennessee 37831, USA}
\affiliation{National Laboratory of Solid State Microstructures, Nanjing University, Nanjing 210093, China}
\author{Qinfang Zhang}
\affiliation{Key Laboratory for Advanced Technology in Environmental Protection of Jiangsu Province, Yancheng Institute of technology, Yancheng 224051, China}
\affiliation{Computational Condensed Matter Physics Laboratory, RIKEN, Wako, Saitama 351-0198, Japan}
\affiliation{CREST, Japan Science and Technology Agency (JST), Kawaguchi, Saitama 332-0012, Japan}
\author{Seiji Yunoki}
\affiliation{Computational Condensed Matter Physics Laboratory, RIKEN, Wako, Saitama 351-0198, Japan}
\affiliation{CREST, Japan Science and Technology Agency (JST), Kawaguchi, Saitama 332-0012, Japan}
\affiliation{Computational Materials Science Research Team, RIKEN AICS, Kobe, Hyogo 650-0047, Japan}
\author{J.-M. Liu}
\affiliation{National Laboratory of Solid State Microstructures, Nanjing University, Nanjing 210093, China}
\affiliation{International Center for Materials Physics, Chinese Academy of Sciences, Shenyang 110016, China}
\author{Elbio Dagotto}
\affiliation{Department of Physics and Astronomy, University of Tennessee, Knoxville, Tennessee 37996, USA}
\affiliation{Materials Science and Technology Division, Oak Ridge National Laboratory, Oak Ridge, Tennessee 37831, USA}
\date{\today}

\begin{abstract}
The two-orbital double-exchange model is employed for the study of
the magnetic and orbital orders in ($R$MnO$_3$)$_n$/($A$MnO$_3$)$_{2n}$
($R$: rare earths; $A$: alkaline earths) superlattices.
The A-type antiferromagnetic
order is observed in a broad region of parameter space
for the case of SrTiO$_3$ as substrate,
in agreement with recent experiments and first-principles calculations
using these superlattices.
In addition, also a C-type antiferromagnetic state
is predicted to be stabilized when using substrates like LaAlO$_3$ with smaller
lattice constants than SrTiO$_3$, again in agreement with first principles results.
The physical mechanism for the stabilization of the A- and C- magnetic transitions
is driven by the orbital splitting of the $x^2-y^2$ and $3z^2-r^2$ orbitals. This
splitting is induced by the $Q_3$ mode of Jahn-Teller distortions created
by the strain induced by
the substrates.  In addition to the special example of
(LaMnO$_3$)$_n$/(SrMnO$_3$)$_{2n}$, our phase diagrams can be valuable for the case
where the superlattices are prepared employing narrow bandwidth manganites.
In particular, several non-homogenous magnetic profiles are predicted to occur
in narrow bandwidth superlattices, highlighting the importance of carrying out
investigations in this mostly unexplored area of research.
\end{abstract}
\maketitle

%
\section{Introduction}

In recent years, remarkable progress has been achieved in the experimental and theoretical
study of complex oxide heterostructures. At present, atomic-scale smooth interfaces
involving pairs of oxides with similar crystal
structures and lattice constants can be routinely
synthesized and characterized with sophisticated techniques, potentially leading to a variety of interesting electronic devices.\cite{Mannhart:Sci,Takagi:Sci,Hammerl:Sci,Hwang:Nm,Dagotto:Sci07,Bibes:Ap,Martin:Mse}
As a prototype of correlated electron system, manganites have been widely involved in complex oxides heterostructures, e.g. manganite-cuprate,\cite{Chakhalian:Sci,Visani:Np} manganite-titanate,\cite{Choi:Prb} manganite-nickelate,\cite{Gibert:Nm} and manganite-BiFeO$_3$.\cite{Wu:Nm10} 
Even pure manganite-manganite superlattices made by antiferromagnetic insulators
LaMnO$_3$ (LMO) and SrMnO$_3$ (SMO) are quite nontrivial.\cite{Koida:Prb,Bhattacharya:Apl,Smadici:Prl,Adamo:Apl,Adamo:Prb,Bhattacharya:Prl,May:Prb,Perucchi:Nl,May:Nm,Galdi:Prb,Yamada:Prb,Monkman:Nm} A metal-insulator transition was observed with increasing $n$ for the case of (LaMnO$_3$)$_{2n}$/(SrMnO$_3$)$_n$ superlattices,\cite{Bhattacharya:Prl,Adamo:Prb,Perucchi:Nl} with LMO to SMO in proportion of $2$ to $1$.
Both employing microscopic models and density-functional theory (DFT) calculations,
the presence of magnetic modulations in these superlattices have been reported,
which may be responsible for the metal-insulator transition.\cite{Dong:Prb08.3,Nanda:Prb09}

Recently, May \textit{et al.} observed an enhanced N\'eel temperature of the
A-type antiferromagnetic (A-AFM) order in the
(LaMnO$_3$)$_n$/(SrMnO$_3$)$_{2n}$
($n$=$1$, $2$) superlattices,\cite{May:Nm}
where now the LMO and SMO are in proportion of $1$ to $2$.
This is surprising since
from the theory perspective it is not straightforward to understand the presence of a
uniform A-AFM order because previous theoretical studies (using both model and DFT approaches) predicted that the ground state spin order within the SrMnO$_3$ region was G-type antiferromagnetic (G-AFM) even when there were only two consecutive SrO sheets as
in the case of (LaMnO$_3$)$_4$/(SrMnO$_3$)$_2$.\cite{Dong:Prb08.3,Nanda:Prb09}
In the (LaMnO$_3$)$_n$/(SrMnO$_3$)$_{2n}$ superlattices,
the consecutive SrO sheets can reach up to $4$ when $n$=$2$.
Thus, the $e_{\rm g}$ electron density deep within this SrMnO$_3$ region
is expected to be very close to zero
as in the case of bulk SrMnO$_3$. For such a low electronic density,
typically the G-AFM order is robust and the A-AFM order would be unusual
according to the corresponding phase diagrams for bulk compounds.\cite{Tokura:Rpp}
However, contrary to these expectations, the experiments
of May \textit{et al.} unveiled a robust A-AFM order
which was uniform {\it all through} the superlattices without any noticeable
magnetic modulation.\cite{May:Nm}
Thus, it is very important to perform better refined
theoretical studies to understand this puzzling behavior.

Very recently, some of the authors performed a DFT calculation that actually
confirmed the presence of a nearly uniform A-AFM phase
for the case of (LaMnO$_3$)$_n$/(SrMnO$_3$)$_{2n}$ when using SrTiO$_3$ (STO) as a substrate
and employing an on-site Hubbard $U$ repulsion not
too large (e.g. $U<1.5$ eV).\cite{Zhang:Prb12} Furthermore,
a uniform C-type antiferromagnetic (C-AFM) order was predicted to occur
for these superlattices for the case of a LaAlO$_3$ (LAO) substrate
also for $U<1.5$ eV.\cite{Zhang:Prb12}
In this previous DFT study, the effect of strain was emphasized
to tune the orbital and magnetic orders. The role of strain to
stabilize A-AFM and C-AFM phases for La$_{1-x}$Sr$_x$MnO$_3$ thin film and (LaMnO$_3$)$_1$/(SrMnO$_3$)$_{1}$ superlattice was also discussed
before from the perspective of DFT calculations.\cite{Fang:Prl,Nanda:Prb}

In the present manuscript, the ($R$MnO$_3$)$_n$/($A$MnO$_3$)$_{2n}$
superlattices will be revisited now using the two-orbital double-exchange
model, as opposed to a first-principles study.
Our mission is not only to confirm (or refute) the DFT predictions
from a conceptually different perspective, but also to reveal
in more detail the physical mechanisms behind the results, particularly
with regards to the relationship between magnetism and orbital population.
Furthermore, our model calculations can go beyond (LaMnO$_3$)$_n$/(SrMnO$_3$)$_{2n}$
and shed light on generic narrow-bandwidth ($R$MnO$_3$)$_n$/($A$MnO$_3$)$_{2n}$
superlattices which have not been experimentally prepared thus far and that are
difficult to analyze accurately with DFT methods. Our study confirms
the presence of the A-AFM and C-AFM states, in good agreement with DFT, but
it has also unveiled
a plethora of other phases, including exotic states with non-uniform configurations,
that potentially can emerge in narrow bandwidth superlattices after the appropriate
selection of elements in the chemical formula and of the substrate.

The organization of the manuscript is as follows. In Section II,
the model and the method used are presented, with
a discussion of the profile of the electrostatic potential
used for the superlattices as well as the variational states employed. The
importance of using twisted boundary conditions, particularly for the A-AFM and C-AFM
phases, is discussed in this section as well.
Section III contains the main results and specific conclusions
for the cases of the two substrates STO and LAO under study here.
Emphasis is given
to phase diagrams varying the superexchange and the electrostatic
potential strength. Also in this section, the
orbital occupation and electronic density for the many layers are
discussed, as well as the implications of our
results for more narrow bandwidth manganites. A summary is provided in Section IV.

%
\section{Model, Method, and Boundary Conditions}

In this investigation, the standard
two-orbital double-exchange model will be employed, in the widely used limit of
an infinite Hund coupling for simplicity. Work carried out over several
decades by dozens of groups has repeatedly shown that this double-exchange
model provides a realistic starting approximation
for manganite compounds.\cite{Dagotto:Prp,Dagotto:Bok,Dagotto:Ssc}
More explicitly, the model Hamiltonian is given by
\begin{eqnarray}
\nonumber H&=&-\sum_{<ij>}^{\alpha\beta}t^{\vec{r}}_{\alpha\beta}(\Omega_{ij}c_{i\alpha}^{\dagger}c_{j\beta}+H.c.)+J_{\rm AFM}\sum_{<ij>}\vec{S}_{i}\cdot \vec{S}_{j}\\
&&+\lambda\sum_{i}Q_{3i}\tau_{zi}+\sum_{i}V_in_i.
\end{eqnarray}

In this Hamiltonian, the first term represents the standard double-exchange hopping process for the
$e_{\rm g}$ electrons. The operators
$c_{i\alpha}$ ($c_{i\alpha}^{\dag}$) annihilate (create) an $e_{\rm g}$ electron
at the orbital $\alpha$ of the lattice site $i$. The spin index is no longer necessary since the spin of
the $e_{\rm g}$ electrons is always
parallel to the
localized $t_{\rm 2g}$ spin $\vec{S}_{i}$ due to the infinite Hund coupling approximation.
The three nearest-neighbor (NN) hopping directions are denoted by $\vec{r}$. The Berry phase
$\Omega_{ij}$ generated by the infinite Hund coupling approximation equals
$\cos(\theta_{i}/2)\cos(\theta_{j}/2)+\sin(\theta_{i}/2)\sin(\theta_{j}/2)\exp[-i(\phi_{i}-\phi_{j})]$, where $\theta$ and $\phi$ are the polar and azimuthal angles
of the $t_{\rm 2g}$ spins, respectively, and $i$ and $j$ are NN sites.
Two $e_{\rm g}$ orbitals ($a$: $d_{x^2-y^2}$ and $b$: $d_{3z^2-r^2}$) are involved in the double-exchange process for manganites, with the hopping amplitudes given by:
\begin{eqnarray}
\nonumber t^x&=&\left(
\begin{array}{cc}
t^x_{aa} &  t^x_{ab} \\
t^x_{ba} &  t^x_{bb}
\end{array}
\right) =\frac{t_0}{4}\left(
\begin{array}{cc}
3 &  -\sqrt{3} \\
-\sqrt{3} &  1
\end{array}
\right),\\
\nonumber t^y&=&\left(
\begin{array}{cc}
t^y_{aa} &  t^y_{ab} \\
t^y_{ba} &  t^y_{bb}
\end{array}
\right) =\frac{t_0}{4}\left(
\begin{array}{cc}
3 &  \sqrt{3} \\
\sqrt{3} &  1
\end{array}
\right),\\
t^z&=&\left(
\begin{array}{cc}
t^z_{aa} &  t^z_{ab} \\
t^z_{ba} &  t^z_{bb}
\end{array}
\right) =t_0\left(
\begin{array}{cc}
0 &  0 \\
0 &  1
\end{array}
\right).
\end{eqnarray}
Below, in the presentation of our results $t_0$ will be considered as the unit of energy.
The second term of the Hamiltonian is the antiferromagnetic superexchange (SE) interaction
between the NN $t_{\rm 2g}$ spins. The typical value of the SE coupling $J_{\rm AFM}$
is approximately between $0.05t_0$ and $0.1t_0$,
based on a variety of previous investigations
for bulk manganites.\cite{Dong:Prb08.3,Dagotto:Prp,Dagotto:Bok,Dagotto:Ssc}
The third term stands for the electron-lattice interaction, with $\lambda$ being
a dimensionless coupling. A crucial aspect of the present investigation is the inclusion of the
strain caused by the substrate. This strain is here modeled
using the $Q_3$ ($\sim(2\delta_z-\delta_x-\delta_y)$) mode of the Jahn-Teller (JT) distortions,
where $\delta_{\alpha}$ stands for the
change in the length of the O-Mn-O bonds along a particular axis $\alpha$. The operator $\tau_{z}$ ($=c_{a}^{\dag}c_{a}-c_{b}^{\dag}c_{b}$) is an orbital pseudospin operator.
The term $\lambda Q_3\tau_z$ splits the energy levels of the orbitals $d_{x^2-y^2}$ and $d_{3z^2-r^2}$.

From a formal perspective, the double-exchange hopping amplitudes, as well as the superexchange and
Jahn-Teller couplings, may be different at the $R$MnO$_3$ and $A$MnO$_3$ regions, and at the interface.
However, since the in-plane lattice constants are fixed to those of the substrate by the growing process,
then the in-plane interaction couplings will be considered almost uniform all through the superlattices.
With regards to the out-of-plane coupling strengths, in principle they should take different values
for $R$MnO$_3$, $A$MnO$_3$, and the interface. However, to fully consider these effects additional
parameters are needed since their precise values are not available in the current state of investigations
in oxide superlattices. Thus, in order to avoid having too many free parameters, it is a practical
choice to carry out these investigations assuming uniform couplings, as used in previous work.(25,34,35)
This approximation reduces the numbers of model parameters needed in our study, allowing us to focus
on the most important aspects of the physics of the problem.\cite{Dong:Prb08.3,Nanda:Prb10,Yu:Prb09} This approximation reduces the numbers of model parameters needed in our study, allowing us to focus on the most important aspects of the physics of the problem. 

In our study neither the breathing
mode $Q_1$ nor the JT mode $Q_2$ are considered in the calculations for the
following reasons. First, the contribution that arises from $Q_1$, which
effectively modifies the on-site potential,
can be simply merged into the fourth term. Second, with regards to the $Q_2$ JT mode
its average value $<Q_2>$ is zero if the substrate is biaxially isotropic.
Although the local values of $Q_{2i}$ may become nonzero in special cases
such as for the undoped manganites $R$MnO$_3$ with orbital order,\cite{Hotta:Prb99}
the $Q_{2i}$ profile is not considered important in the bulk doped
manganites such as $R_{1/3}A_{2/3}$MnO$_3$~\cite{Hemberger:Prb02}
and, more importantly for our studies on superlattices,
they have not been reported as being of relevance in previous studies of
($R$MnO$_3$)$_n$/($A$MnO$_3$)$_{2n}$ superlattices to our knowledge.
Therefore, the $Q_{2i}$ term will be neglected in the present effort for mere simplicity.

In our calculations,  $Q_{3i}$ is proportional to the local quantity ($c/a-1$),
where $c$ and $a$ are the out-of-plane and in-plane lattice constants, respectively.
The specific generic values of $Q_{3i}$ used in the present work are listed in Table~\ref{table}.
In the present effort, the superlattices are assumed to be fully strained following the
underlying substrates, as observed experimentally.\cite{May:Nm}

\begin{figure}
\centering
\includegraphics[width=0.5\textwidth]{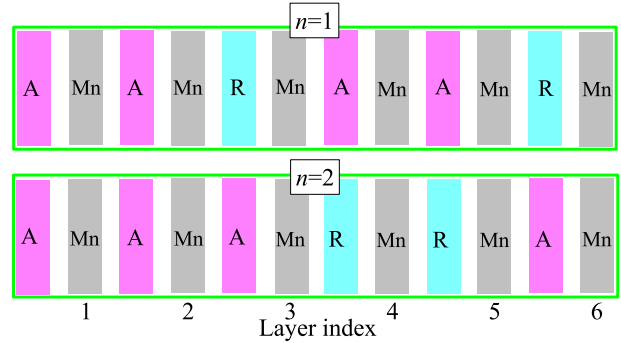}
\caption{(Color online) Sketch showing the
distribution of layers in the finite cluster with six Mn-oxide layers
used in our studies. ``Mn'' stands for the MnO$_2$ layers,
``R'' for the $R$O layers,
and ``A'' for the $A$O layers. The only active layers for the mobile electrons
are the Mn layers in our study, but
the other layers influence on the Mn layers by
their electrostatic potential, as discussed in the text and shown in Table I.
}
\label{sketch}
\end{figure}

The fourth term is the on-site Coulomb potential energy, and $n_i$ is the local $e_{\rm g}$
electronic density. $V_i$ is the Coulomb potential arising from the A-site cations, from the $e_{\rm g}$ electrons,
and from the oxygen anions (i.e. the $Q_1$ mode distortion mentioned before).
In the rest of the publication, the profile of $V_i$ will be approximately
determined by the NN A-site cations,\cite{Bouzerar:Prb,Dong:Prb08.3} as shown in Table~\ref{table}, while
its actual magnitude $V$
will be considered a parameter and phase diagrams will be constructed varying this parameter.
Fixing such a profile is not a drastic approximation since previous more detailed studies,
even using the self-consistent Poisson equation, have
shown that these sophisticated
approaches tend to produce smooth potentials with profiles
similar to those used here.\cite{Yu:Prb09}
For the benefit of the reader, the distribution of the MnO$_2$, $R$O, and $A$O
layers present in our superlattices are shown in Fig.~\ref{sketch}.

The above described model Hamiltonian has been numerically solved here
using finite clusters. To simulate the case of the
($R$MnO$_3$)$_n$/($A$MnO$_3$)$_{2n}$ ($n$=$1$, $2$)
superlattices, a $4\times4\times6$ cluster
with twisted boundary conditions (TBC)\cite{Salafranca:Prb09} is here adopted. The TBC with proper
$k$-meshes ($k\times k\times k$ used here) are often employed
to reach larger lattices, thus
reducing the finite-size lattice effects (which are indeed of relevance
in the present study, as shown below).\cite{Dong:Prb11}

\begin{figure}
\vskip -0.5cm
\centering
\includegraphics[width=0.45\textwidth]{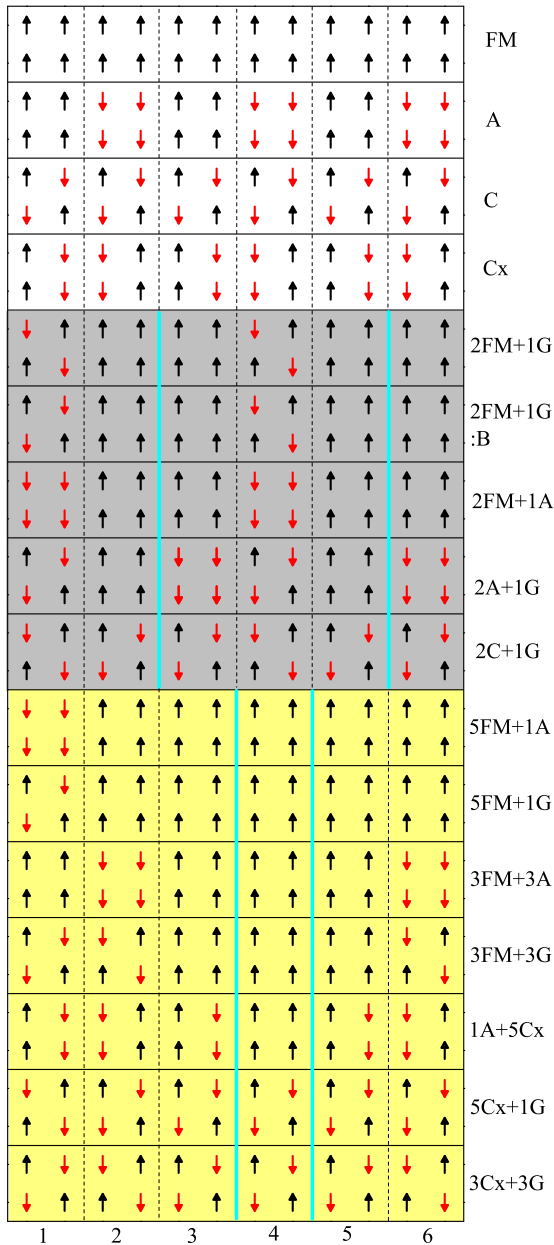}
\vskip -0.5cm
\caption{(Color online) Sketch of the spin patterns
for several of the variational states considered here.
The index for the six layers of MnO$_2$
in our finite-cluster study is shown at the bottom.
In this figure, only a $2\times2$ unit for each layer is shown. The rest of the magnetic
state at each layer is obtained by repeating the pattern provided.
The first four configurations (i.e. those with the white background) are uniform
and common for the $n$=$1$ and $n$=$2$ cases. States of this kind also appear in the
bulk limit and are not affected by the superlattice periodicity.
The next five configurations shown (with the
gray background) are only for the $n$=$1$ case. The last
seven configurations (with the yellow background) are only for the $n$=$2$ case.
In the non-uniform cases, the cyan bars indicate the location of
the $R$O sheets (while the $A$O sheets are not shown explicitly).
The non-uniform phases are denoted using a notation $mX+m'Y$, where $X$ and $Y$
are abbreviations for the $X$-AFM and $Y$-AFM phases, and $m$/$m'$
are the fractions of the $X$/$Y$ phases, respectively.}
\label{pattern}
\end{figure}

To determine the ground state phases in
these superlattices, a variational procedure is used, namely
the energies of candidate phases are compared in the $J_{\rm AFM}$-$V$ parameter space,
for several values of $\lambda Q_3$ as an additional parameter.
The candidate phases are proposed according
to the symmetries of the superlattices. Contrary to other variational studies limited to
a relatively small number of proposed states, in this effort $30$ candidates
for the $n$=$1$ case and $56$ candidates for the $n$=$2$ case have been considered.
These candidate states were constructed
by combining the well-known FM, A-AFM, C-AFM, and G-AFM spin
arrangements in each of the layers (note that the CE-AFM state is not considered
since the $Q_2$ mode is neglected).
Some of these candidates will appear explicitly in the
phase diagrams described below in Section III, and some of them are presented
in Fig.~\ref{pattern} for the benefit of the reader.

\begin{table}
\caption{Parameter profiles used in the present simulations. Upper rows:
the value of $Q_{3i}$s in the three regions of relevance, for the two substrates considered here,
namely for MnO$_2$ layers in between two $R$O layers (represented as $R$MnO$_3$),
in between a $R$O layer and a $A$O layer (represented as $R_{1/2}A_{1/2}$MnO$_3$),
and finally in between two $A$O layers (represented as $A$MnO$_3$).
The lattice constants of $R$MnO$_3$ and SrTiO$_3$ are similar,
while those of $A$MnO$3$ and LaAlO$_3$ are also similar.
Here $Q$ is positive. Lower row: value of $V_i$ in the three regions.}
\begin{tabular*}{0.5\textwidth}{@{\extracolsep{\fill}}lllllr}
\hline \hline
$Q_{3i}$ & $R$MnO$_3$ & $R_{1/2}A_{1/2}$MnO$_3$ & $A$MnO$_3$ \\
\hline
SrTiO$_3$ & $0$ & $-Q/2$ & $-Q$\\
LaAlO$_3$ & $Q$ & $~~~Q/2$ & $~~~0$\\
\hline \hline
$V_i$ & $0$ & $~~~V/2$ & $~~~V$\\
\hline \hline
\end{tabular*}
\label{table}
\end{table}

%

\subsection{Discussion of finite-size lattice effects}

In several previous computational studies of the double-exchange model,
periodic boundary conditions (PBC) were used for Monte Carlo simulations
carried out in real-space lattices.\cite{Dong:Prb08.3}
Due to the substantial CPU-time consumed by the frequent
diagonalizations in the fermionic sector,
calculations have been typically limited to relatively small lattice sizes,
especially for three-dimensional studies. For example, in several cases
the in-plane lattice cell was only $4\times4$ in size,\cite{Dong:Prb08,Dong:Prb08.3,Yu:Prb09}
which might give rise to non-negligible
finite-size effects. These effects are not very
prominent for the ferromagnetic (FM), G-AFM, or even the CE-type antiferromagnetic cases,
which were analyzed in previous studies.\cite{Dong:Prb08,Dong:Prb08.3,Yu:Prb09}
However, as it will be described below, in the effort described here
the dominant states involved are the A-AFM and C-AFM phases in various combinations.
It is necessary to verify that the finite-size lattice effects are not too large
for these phases. Moreover, the presence of a superlattice structure may increase the
relevance of these effects.

\begin{figure}
\includegraphics[width=0.49\textwidth]{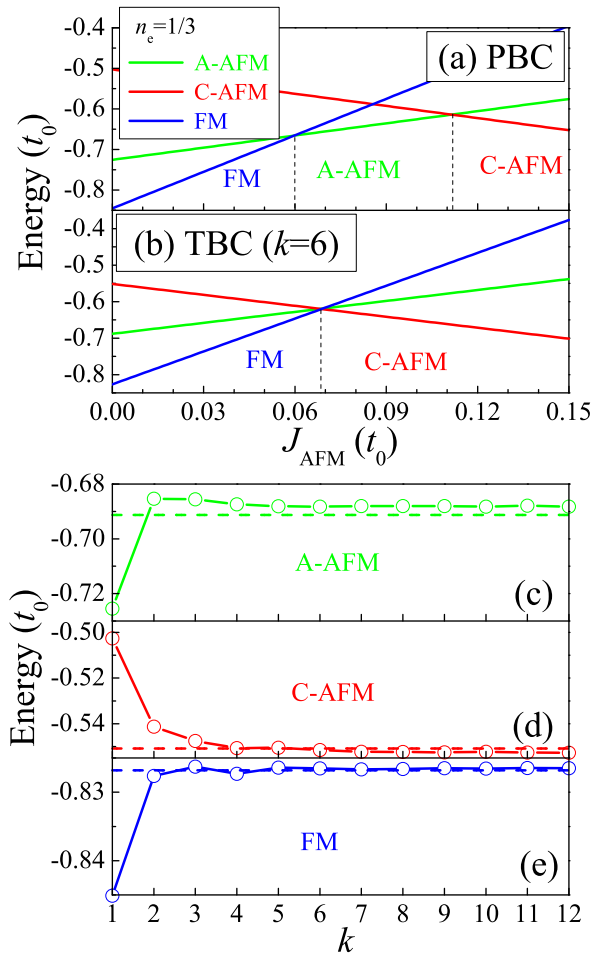}
\caption{(Color online) A comparison of energies using
a $4\times4\times6$ lattice (without including the JT distortion $Q_{3i}$
and the Coulomb potential $V_i$, for simplicity) contrasting the cases of
the PBC and the TBC. The overall $e_{\rm g}$-electron density is chosen as $n_e$=$1/3$.
Panels (a-b) show the total energies (i.e. double-exchange and superexchange)
per site as a function of the superexchange coupling. (a) is for PBC, while
(b) is for TBC with $k$=$6$. The ``coincidence'' of the
three energies in (b) at one particular value of $J_{\rm AFM}$ is accidental.
(c-e) Double-exchange energy for each of the cases indicated, as a function of $k$ in the
TBC mesh.
The accurate energies for the bulk limit obtained analytically~\cite{Dong:Jpcm}
are also shown as dashed lines to guide the eye. It is obvious that the use of
TBC can drastically reduce the finite-size effects.}
\label{k}
\end{figure}

This present study of size effects and boundary conditions
will be carried out first without
the superlattice structure, namely mimicking the bulk system. The conclusions of this
subsection will then be applied to the superlattice study in the rest of the manuscript.
In Figs.~\ref{k}~(a-b), the ground state energies (per site) of the
FM, A-AFM, and C-AFM phases are calculated for the case of bulk
$R_{1/3}A_{2/3}$MnO$_3$ using the above described Hamiltonian
(and with $Q_{3i}$=$0$ and $V_i$=$0$, for simplicity, i.e. no superlattice is considered in this test,
 as already explained).
The energies of these three states are compared for both the cases of PBC and TBC.
In the PBC case, the ground state changes from FM to A-AFM, and then to C-AFM
with increasing superexchange $J_{\rm AFM}$ (or equivalently by decreasing
the overall bandwidth). As shown in Fig.~\ref{k}(a), with these boundary conditions the A-AFM state
is stable from $J_{\rm AFM}=0.06$ to $J_{\rm AFM}=0.11$.
However, the A-AFM state is actually unstable with the use of the TBC,
and the ground state changes from FM to C-AFM directly, as shown in Fig.~\ref{k}(b), with
the A-AFM coexisting with the other two states at just one point.
Therefore, to reduce this finite-size lattice effect, the TBC
will be adopted in the rest of the calculations described here.
However, note that the use of
TBC considerably increases the CPU-time of the study,
which renders a fully unbiased Monte Carlo simulation too time consuming. However,
our use of dozens of variational states give us confidence that the present results have
captured the essence of the problem.

In Figs.~\ref{k}(c-d), the double-exchange energies (per site) are shown
as a function of the value of $k$ in the TBC.
From the energy values in the vertical axes, it is clear
that the per-site energy deviation from the bulk value using the PBC
(corresponding to $k$=$1$) is the most serious for the case of the
C-AFM state (up to $-0.05$$t_0$), milder for the A-AFM state (up to $0.037$$t_0$),
and the less important for the FM state (difference less than $0.019$$t_0$).
And, moreover, $k$=$6$ or even less in the TBC
is certainly enough for the energy per site to overcome
those finite-size lattice effects. Thus, this value of $k$
will be adopted in the calculations for the superlattices presented below.

%

\section{Results and Discussion}

\subsection{($R$MnO$_3$)$_n$/($A$MnO$_3$)$_{2n}$ on SrTiO$_3$}


As explained in the Introduction,
the results of our investigations will be provided for two particular substrates, starting
with SrTiO$_3$. For this case, note that the average lattice constant for a
pseudo-cubic $R$MnO$_3$ ($R$=La, Pr, Nd) is close to the lattice
constant of SrTiO$_3$ ($3.905$ \AA). More specifically, this (average) lattice constant
is about $3.94$ \AA{} for $R$=La, $3.916$ \AA{} for $R$=Pr, and $3.907$ \AA{}
for $R$=Nd.\cite{Alonso:Ic} Then, in the numerical simulations to describe this particular
case, the $R$MnO$_3$ layers grown on SrTiO$_3$ are assumed to be nearly cubic,
namely the value $Q_{3i}=0$ is used for the active JT mode in our study.
In contrast, the cubic $A$MnO$_3$ has a much smaller lattice constant,
in particular $3.805$ \AA{} for $A$=Sr and $3.727$ \AA{}
for $A$=Ca,\cite{Chmaissem:Prb} which gives rise to an important in-plane expansion
and out-of-plane reduction of the lattice constants when growing
on the SrTiO$_3$ substrate. The ratio $c/a$ 
of $A$MnO$_3$ on SrTiO$_3$ will be smaller
than $1$, implying a negative $Q_{3i}$. Therefore, in the following
simulations with SrTiO$_3$ as the substrate, the values of the $Q_{3i}$ variables are
considered to be $0$ in the $R$MnO$_3$ region (more specifically, when a MnO$_2$
layer is in between two $R$O layers), but $-Q$ in the $A$MnO$_3$ region (i.e. when the
MnO$_2$ layer is in between two $A$O layers),
where $Q$ is a positive parameter to be tuned in our study.
For the interfacial $R$O-MnO$_2$-$A$O layer,
an average value $-Q/2$ is used as a first order approximation,
as shown in Table~\ref{table}.

The DFT results\cite{Nanda:Prb09,Zhang:Prb12}
on (LaMnO$_3$)$_{2n}$/(SrMnO$_3$)$_n$
and (LaMnO$_3$)$_{n}$/(SrMnO$_3$)$_{2n}$ superlattices
($n$=$1$,$2$) have indicated that  the Coulomb energy difference
between LaMnO$_3$ and SrMnO$_3$ is about $0.7$-$1$ eV, which
corresponds to $1.4$-$2$$t_0$ in our simulation if $t_0$ is
$0.5$ eV (a reasonable approximate value for wide
bandwidth manganites).\cite{Dong:Prb08.3} However, besides
the Coulomb potential from the A-site cations, the $V_i$ used
here also contains other contributions, such as those  from the breathing
phononic $Q_1$ mode. In fact, previous model studies
found that $V_i$=$0.6$-$0.9$$t_0$ is a proper range for this potential in
the double-exchange model.\cite{Dong:Prb08.3} However, for completeness
here the full range of $V$ up to $2$$t_0$ will be investigated in the phase
diagrams.

\begin{figure}
\centering
\includegraphics[width=0.6\textwidth]{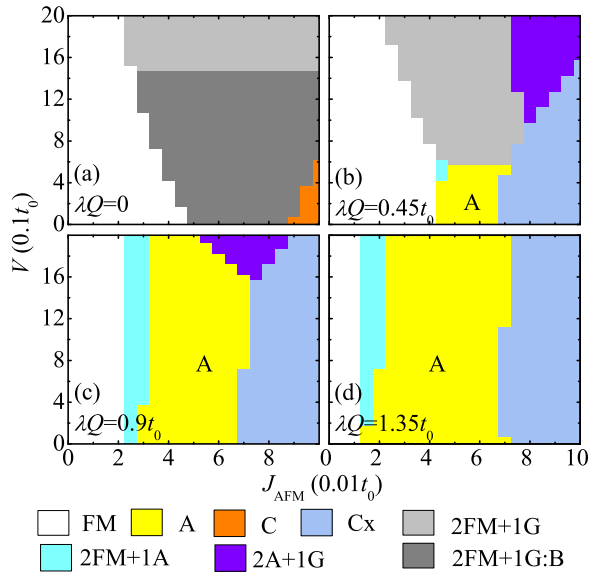}
\caption{(Color online) Phase diagram for the case of ($R$MnO$_3$)$_1$/($A$MnO$_3$)$_2$
with SrTiO$_3$ as substrate, at several values of $\lambda Q$s.}
\label{l1s2sto}
\end{figure}

The ground state phase diagrams corresponding to ($R$MnO$_3$)$_1$/($A$MnO$_3$)$_2$ ($n$=$1$)
using SrTiO$_3$ as the substrate are shown in Fig.~\ref{l1s2sto} at various values of $\lambda Q$.
Without the JT distortion (i.e. for $\lambda Q=0$, see Fig.~\ref{l1s2sto}~(a)),
the A-AFM phase is {\it not} stable in this superlattice despite varying the
parameters $J_{\rm AFM}$ and $V$. Instead, FM tendencies dominate with at most just $1/3$
of the superlattice being G-AFM even at the largest $J_{\rm AFM}$ studied here. The
presence of one G-AFM layer for every two FM layers is in agreement with the naive
intuition described in the Introduction, namely that the G-AFM state
should be favored inside $A$MnO$_3$.
However, these results change qualitatively when the strain effects are
incorporated. In fact,
with increasing $\lambda Q$, the A-AFM phase now emerges
around $J_{\rm AFM}$=$0.05$-$0.06$$t_0$ and takes over an increasing fraction
of parameter space with increasing $\lambda Q$. This effect is the most prominent
in the absence of the electrostatic potential.
In principle, a strong Coulomb potential $V$, which reduces the $e_{\rm g}$ electronic density
in the $A$MnO$_3$ region, can
suppress this uniform A-AFM order. In fact, in Fig.~\ref{l1s2sto}~(b) the A-AFM phase
is suppressed rapidly with increasing $V$.
However, in spite of this negative influence of $V$,
when the JT distortion is strong enough, e.g. $\lambda Q\geqslant0.9t_0$,
the A-AFM phase is stable and quite robust in the broad region of parameter space
investigated here, in agreement with the experiments
described in the Introduction.
Note also the appearance of additional robust
states with further increasing $J_{\rm AFM}$,
particularly the Cx state.
As discussed later in the manuscript, these states
could appear in manganites with a bandwidth smaller than for the case $A$=Sr.

\begin{figure}
\centering
\includegraphics[width=0.6\textwidth]{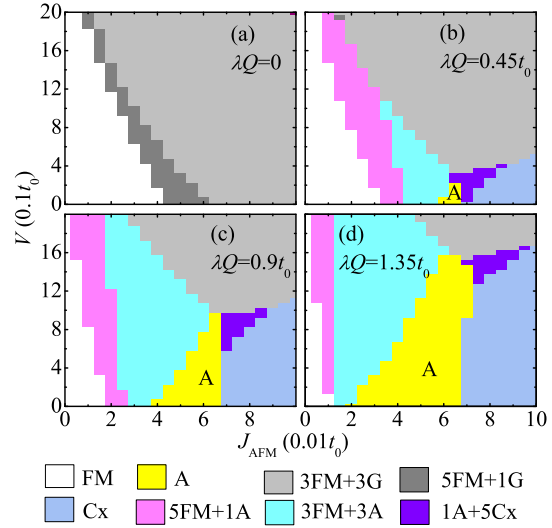}
\caption{(Color online) Phase diagram for the case of
($R$MnO$_3$)$_2$/($A$MnO$_3$)$_4$ with SrTiO$_3$ as substrate, at several
values of $\lambda Q$.}
\label{l2s4sto}
\end{figure}

For the case of the ($R$MnO$_3$)$_2$/($A$MnO$_3$)$_4$ superlattice
with SrTiO$_3$ as substrate, the strong JT distortion is even more
crucial to obtain the A-AFM phase, as shown in Fig.~\ref{l2s4sto}.
The A-AFM phase is stable only  in the interval $J_{\rm AFM}$=$0.05$-$0.06$$t_0$
for the cases of a  strong JT distortion and weak to intermediate Coulomb potentials.
Comparing with the previously described case of $n=1$,
the A-AFM phase appears to be more fragile in the $n=2$ case,
with stability regions in parameter space that are smaller in size for $n$=2
than for $n$=1.
This tendency is in agreement with the experimental results
reporting that the A-AFM N\'eel temperature of (LaMnO$_3$)$_1$/(SrMnO$_3$)$_2$
is higher than (LaMnO$_3$)$_2$/(SrMnO$_3$)$_4$.\cite{May:Nm}
The physical reason is that the charge density deep into the $A$MnO$_3$
region is very low due to the Coulomb potential $V$. Thus, the combination
of $3$ FM layers ($R$MnO$_3$ and interfacial layers) plus $3$ G-AFM layers
($A$MnO$_3$ layers) eventually wins in the energy competition when $V$ is large enough,
in agreement with the naive expectations presented in the Introduction. However,
at intermediate values of $J_{\rm AFM}$ and for large enough strain, represented
by $\lambda Q$, the experimentally observed A-AFM state is stabilized.

Increasing further $J_{\rm AFM}$, the ground state becomes a Cx-type
antiferromagnetic (Cx-AFM) phase.
Here, the Cx-AFM is similar to the C-AFM phase, but the chains with the spins aligned
lay in the $a-b$ plane
while in the standard C-AFM state this spin alignment is along the $c$ axis (out-of-plane).
This phase is certainly interesting but our description
of results here and in the rest of the publication will focus
mainly on the phases that appear at intermediate values of $J_{\rm AFM}$. The
reason is that for the case of bulk manganites, phases not considered in our effort
such as the CE-AFM state
with $Q_2$ distortions
also appear in the range of $J_{\rm AFM}$ studied here at least
at particular electronic densities
such as half-doping. Moreover, other states stabilized
at larger values of $J_{\rm AFM}$ and the same doping, such as the G-AFM state,
have not been found experimentally for bulk compounds.
Thus, it is natural that the focus of our results for superlattices
should only be on the intermediate $J_{\rm AFM}$ range. Nevertheless, a brief discussion
for narrow bandwidth manganites is presented below in this manuscript as well.

%
\subsection{($R$MnO$_3$)$_n$/($A$MnO$_3$)$_{2n}$ on LaAlO$_3$}

Consider now the case of LaAlO$_3$ as the substrate.
In a recent first-principles calculation, the (LaMnO$_3$)$_n$/(SrMnO$_3$)$_{2n}$
superlattices grown on LaAlO$_3$ as the substrate were also studied,
and results quite different from those of the SrTiO$_3$ substrate were obtained,
as already mentioned in the Introduction.\cite{Zhang:Prb12}
For this reason, here the phase diagrams for superlattices grown on LaAlO$_3$ are
also calculated to contrast model Hamiltonian
results against first-principles approaches.

\begin{figure}
\includegraphics[width=0.5\textwidth]{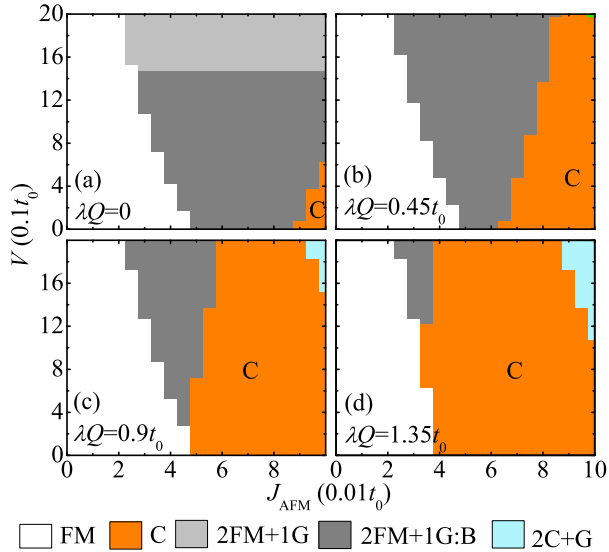}
\caption{(Color online) Phase diagram for the case of
($R$MnO$_3$)$_1$/($A$MnO$_3$)$_2$ with LaAlO$_3$ as substrate, at several
values of $\lambda Q$.}
\label{l1s2lao}
\end{figure}

In contrast to the case of SrTiO$_3$ as substrate,
the lattice constant of LaAlO$_3$ (which is approximately $3.791$ \AA)
is much smaller than the lattice constant of $R$MnO$_3$,
but it is close to SrMnO$_3$ ($3.805$ \AA) and to
CaMnO$_3$ ($3.727$ \AA). Therefore, the $R$MnO$_3$ layers
are considerably compressed in-plane while the $A$MnO$_3$ layers
are almost free from strain when LaAlO$_3$ is the substrate.
Then, to simulate this strain effect
a positive value of $Q_{3i}=Q$ is adopted for the $R$MnO$_3$ layers
while $Q_{3i}=0$ is used in the $A$MnO$_3$ region. Similarly as in the
previous subsection,
for the interfacial layers $Q_{3i}$ is taken as $Q/2$
as a first order approximation, as shown in Table~\ref{table}.
This simple and smooth JT profile is the opposite to that used before for the SrTiO$_3$ case.

As shown in Fig.~\ref{l1s2lao}, the A-AFM phase, which is robust in the
phase diagrams for the  case of SrTiO$_3$ as substrate, no longer
survives for the case of LaAlO$_3$. Instead, now the C-AFM state becomes
the most likely phase appearing in reasonable regions of parameter space
for the large bandwidth manganites. This result agrees with the first-principles prediction
that the C-AFM phase is quite robust in the (LaMnO$_3$)/(SrMnO$_3$)$_2$
superlattice on LaAlO$_3$.\cite{Zhang:Prb12} Note that
in Fig.~\ref{l1s2lao} strain is
crucial to stabilize the C-AFM phase.
Without strain, the naive expectation described
in the Introduction of finding G-AFM layers
in the $A$MnO$_3$ region is indeed fulfill
since the 2FM+1G and 2FM+1G:B states dominate.
Once again, the crucial role of strain
in these investigations becomes clear.

\begin{figure}
\includegraphics[width=0.5\textwidth]{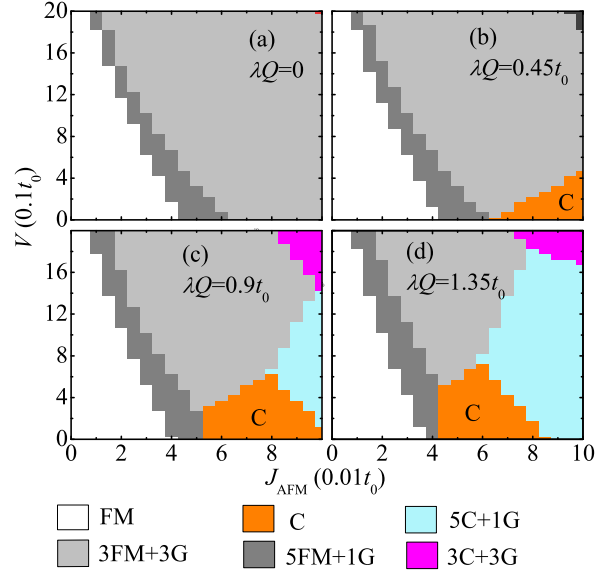}
\caption{(Color online) Phase diagram for the case of ($R$MnO$_3$)$_2$/($A$MnO$_3$)$_4$
with LaAlO$_3$ as substrate, at several values of $\lambda Q$.}
\label{l2s4lao}
\end{figure}

Note, however, that for ($R$MnO$_3$)$_2$/($A$MnO$_3$)$_4$
the uniform C-AFM phase appears to be fragile, and becomes stable
only in a small region varying $V$ (Fig.~\ref{l2s4lao}).
Instead, the hybrid configuration consisting of
$3$ FM layers ($R$MnO$_3$ and interfacial layers) plus $3$ G-AFM layers ($A$MnO$_3$ layers)
is mainly stabilized when $V$ increases. Thus, the experimental realization of the C-AFM
state for $n$=2 using LAO as substrate appears more problematic than for $n$=$1$.
Further increasing $n$, eventually the
naive expectation of a G-AFM dominance in the
$A$MnO$_3$ region must be satisfied, but it is difficult to predict
the precise value of $n$ for the transition
from a C-AFM dominated to a G-AFM dominated behavior in the $A$MnO$_3$ layers.

%
\subsection{Electronic density and orbital occupation}

The $e_{\rm g}$ electronic density and the orbital occupancies
are important factors to understand intuitively the origin
of the previously described ground states of
the ($R$MnO$_3$)$_n$/($A$MnO$_3$)$_{2n}$ superlattices on different substrates.
These quantities are closely related to both the uniform $Q_3$ JT distortions
(determined by the substrates) and the spin arrangements.

\begin{figure}
\centering
\includegraphics[width=0.5\textwidth]{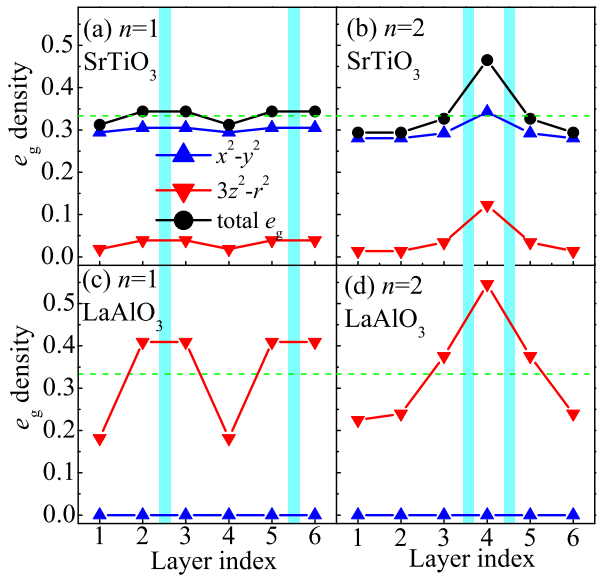}
\caption{(Color online)
The $e_{\rm g}$ density profiles
of the orbital levels in ($R$MnO$_3$)$_n$/($A$MnO$_3$)$_{2n}$
for the special example of  $V$=$0.6t_0$ and $\lambda Q$=$0.9t_0$.
In all panels, the values of $n$ and the substrate used are indicated.
The total $e_{\rm g}$ density (black) and the orbital
occupations (blue: $d_{x^2-y^2}$; red: $d_{3z^2-r^2}$) are shown.
In (c) and (d), the $d_{3z^2-r^2}$ orbital is virtually
the only occupied orbital, and for this reason the total density
is not shown since it is virtually identical to the $d_{3z^2-r^2}$ density.
The cyan bars denote the $R$O sheets in the superlattices.}
\label{Graph9}
\end{figure}

As a typical example, the case of $V$=$0.6t_0$ and $\lambda Q$=$0.9t_0$
will be discussed in detail, and the $e_{\rm g}$ electronic
densities and their orbital components will be calculated explicitly.
As shown in Figs.~\ref{Graph9}~(a) and (b),
for the case of the A-AFM phase on SrTiO$_3$
the $e_{\rm g}$ densities only weakly fluctuate around
the average value $n_e$=$1/3$ all through the six layers of the superlattice
studied here, despite the presence of the electrostatic
potential modulation. In fact, even the lowest $e_{\rm g}$
density within the $A$MnO$_3$ region
is still quite robust and only slightly smaller than $0.3$,
i.e. it is appreciably higher than the
$e_{\rm g}$ electronic density of $A$MnO$_3$ in the bulk (i.e. $\sim0$).
In other words, the transfer of charge from $R$MnO$_3$ to $A$MnO$_3$ in these superlattices
with thin components is quite prominent and important to stabilize the particular spin orders
that have been discussed in this publication. Therefore, the magnetic orders found in the
present manganite superlattices cannot be simply extended to pure $R$MnO$_3$ and $A$MnO$_3$ films
even under the same strain conditions.

This is a crucial observation to understand the stabilization of
the A-AFM order in the superlattice as compared to the G-AFM order in the bulk
of $A$MnO$_3$. Note that in previous
theoretical studies of (LaMnO$_3$)$_{2n}$/(SrMnO$_3$)$_{n}$
superlattices,\cite{Dong:Prb08.3,Nanda:Prb09}
the G-AFM order was found to be stable in the SrMnO$_3$ region
because its local electron density
was much lower than found in the example studied here. The reason for this
difference is the inclusion of strain effects (i.e. a nonzero $\lambda Q$)
in the present analysis.
In fact, the effort described here suggests
that it may be necessary to revisit the previously studied
(LaMnO$_3$)$_{2n}$/(SrMnO$_3$)$_{n}$ system, because the effects of strain were not
taken into account in those previous theoretical investigations.

For the case of the C-AFM phases on LaAlO$_3$ the
situation is different.
In this case, the orbital occupation is purely $d_{3z^2-r^2}$,
without an appreciable $d_{x^2-y^2}$ component.
The $e_{\rm g}$ electronic densities are also substantially
modulated, following much closer
the cations than for the case of SrTiO$_3$.
This charge modulation is prominent for the C-AFM phases
on LaAlO$_3$, while for the A-AFM phases on SrTiO$_3$ the electronic density
is almost uniform for $n$=1 and only weakly modulated for $n$=2,
as it is clear by comparing Figs.~\ref{Graph9}~(a)-(b)
with (c)-(d).

\begin{figure}
\centering
\includegraphics[width=0.43\textwidth]{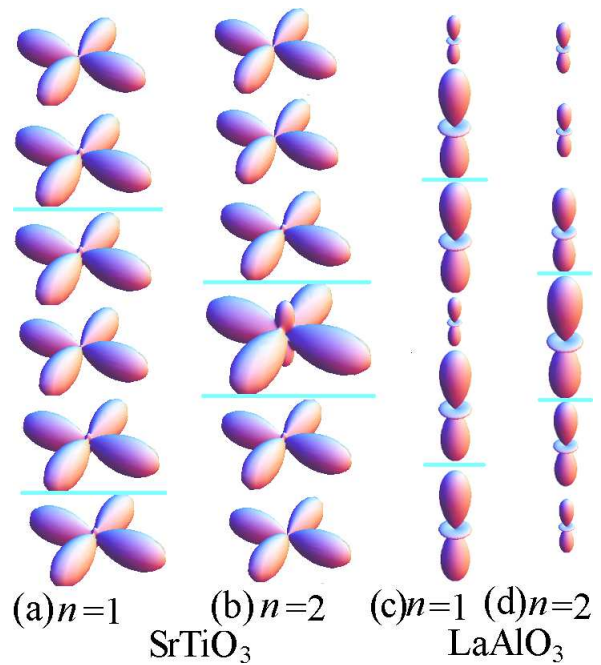}
\caption{(Color online)
Sketch of the dominant orbital occupation for the two substrates and
the two values of $n$ studied in this manuscript. The six drawings for each
case correspond to the six layers studied here, and their size is proportional
to the actual occupation of the respective orbitals.
The lines in cyan color indicate where the $R$O layers are. Since the associated
A-AFM and C-AFM magnetic orders are uniform in the layers that form the superlattice,
then the results in this figure are the same for all the other chains running
in the direction perpendicular to the interfaces.
}
\label{newoo}
\end{figure}

Addressing now the orbital occupation,
for the A-AFM phase on SrTiO$_3$ the
$d_{x^2-y^2}$ is the orbital the most populated,
with  a mild contribution of the $d_{3z^2-r^2}$ component
in the $R$MnO$_3$ portion of the superlattice for $n$=2,
as shown in Figs.~\ref{newoo}~(a,b).
On the other hand, for the C-AFM phase on LaAlO$_3$, it is
the $d_{3z^2-r^2}$ orbital that dominates, as intuitively expected
(see Figs.~\ref{newoo}~(c,d)).
Figs.~\ref{newoo}~(a-d) show only one line
of orbitals running perpendicularly to the interfaces
because the orbital occupation is the same
in all the sites of the $x$-$y$ layers since
the A-AFM and C-AFM are uniform in those layers.

\begin{figure}
\centering
\includegraphics[width=0.5\textwidth]{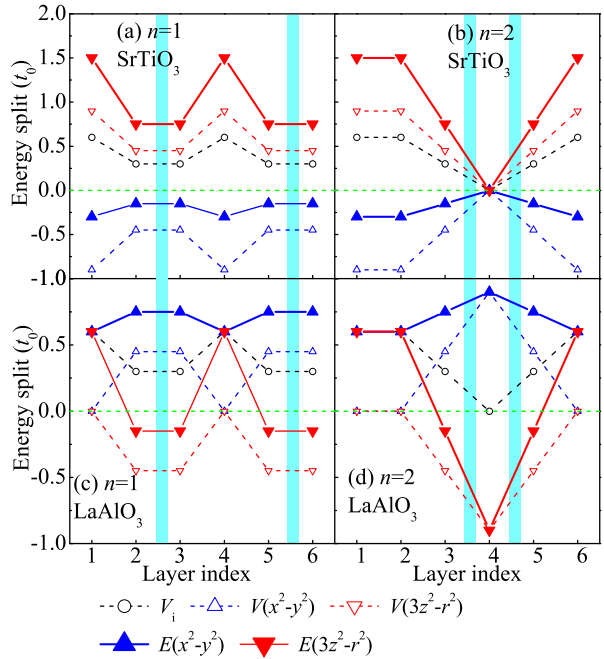}
\caption{(Color online)
The $e_{\rm g}$-orbitals energy splitting and the electrostatic potentials
in ($R$MnO$_3$)$_n$/($A$MnO$_3$)$_{2n}$
for the special example of  $V$=$0.6t_0$ and $\lambda Q$=$0.9t_0$.
In all panels, the values of $n$ and the substrate used are indicated.
The final energy levels of the two orbitals (after the JT splitting plus
the shift from $V_i$) are shown as the blue ($d_{x^2-y^2}$)
and red ($d_{3z^2-r^2}$) solid curves.
The cyan bars denote the $R$O sheets in the superlattices.
The electrostatic potentials $V_i$ are shown
in dashed black curves. For completeness, the energy
split of the two orbitals only due to
the JT $Q_3$ mode is also shown, with the
blue (red) dashed curve denoting the
$d_{x^2-y^2}$ ($d_{3z^2-r^2}$) orbital.
}
\label{Graph10}
\end{figure}

To understand better the density and orbital profiles found in our
variational study, the splitting of the energy
between the two $e_{\rm g}$-orbitals induced by the $Q_3$ mode
is shown in Figs.~\ref{Graph10}~(a-d).
On SrTiO$_3$, the $d_{x^2-y^2}$ and $d_{3z^2-r^2}$
orbital levels in the $A$MnO$_3$ region of the superlattice
are shifted down and up, respectively,
due to the influence of the $\lambda Q_{3i}$ coupling in that region,
while they are degenerate for $n$=2
in the $R$MnO$_3$ portion of the superlattice
due to its nearly cubic lattice symmetry
($\lambda Q_{3i}$=$0$ there). Note that for $n$=1, $\lambda Q_{3i}$ is
always nonzero, thus there is always a splitting between the two orbitals.
Therefore,
the $d_{x^2-y^2}$ level in the $A$MnO$_3$ region is low enough
to ``accumulate'' $e_{\rm g}$ electrons
even when the electrostatic potential $V_i$ is imposed,
as shown in Figs.~\ref{Graph10}~(a,b).
Since the $d_{x^2-y^2}$ orbital, with its in-plane lobes, prefers the in-plane
double-exchange hopping between NN sites,  then the A-AFM state is stabilized
since in this case
the double-exchange process is precisely restricted to occur within the $x$-$y$ plane.

Contrary to the case of SrTiO$_3$,
when LaAlO$_3$ is the substrate the $d_{x^2-y^2}$ and $d_{3z^2-r^2}$ orbital levels
in the $R$MnO$_3$ portion of the superlattice are shifted up and down, respectively,
due to the influence of $\lambda Q_{3i}$,
while they are degenerate in the $A$MnO$_3$ component of the superlattice
 where $\lambda Q_{3i}$=$0$.
Since the double-exchange process in the C-AFM phase is
only active along the out-of-plane  chains with the same orientation of the spin
and since in this direction the hoppings associated with the $x^2-y^2$ vanish
($t_{aa}^z$=$t_{ba}^z$=$t_{ab}^z$=$0$), then
the $d_{3z^2-r^2}$ orbital is occupied
while the $d_{x^2-y^2}$ orbital is virtually empty.

In summary, the two $e_{\rm g}$ orbitals are both active in unstrained layers
but typically only one dominates in the uniformly strained layers.
This simple observation is sufficient to understand several
features of the phase diagrams that were obtained numerically.

%
\subsection{Extensions to narrow-bandwidth manganite superlattices}

Most experiments on Mn-oxide superlattices have focused
on the wide bandwidth manganites involving
LaMnO$_3$ and SrMnO$_3$ as components.\cite{Koida:Prb,Bhattacharya:Apl,Smadici:Prl,Adamo:Apl,Adamo:Prb,Bhattacharya:Prl,May:Prb,May:Nm,Monkman:Nm} For this reason considerable theoretical
investigations have been carried out with this focus
as well.\cite{Dong:Prb08.3,Nanda:Prb09,Zhang:Prb12}
However, some recent theoretical
efforts pointed out that exotic phenomena might exist
at interfaces involving narrow-bandwidth manganites, even including the presence
of novel magnetic states that do not appear in the bulk phase diagrams
of the individual materials that form the superlattice.\cite{Yu:Prb09}
Since reducing the bandwidth is equivalent to increasing
the values of $J_{\rm AFM}$ and $V$, because they are in units of the hopping $t_0$,
then the previously discussed phase diagrams contained in
Figs.~\ref{l1s2sto}-\ref{l2s4lao} indicate that
other magnetic orders beyond the uniform A-AFM and predicted C-AFM states
could be stabilized in these superlattices if the parameters $J_{\rm AFM}$
or $V$ are increased further by using manganites different from
LaMnO$_3$ or SrMnO$_3$.

For example, considering the case of the SrTiO$_3$ substrate,
the uniform Cx-AFM order is neighboring the A-AFM order in the phase
diagrams Figs.~\ref{l1s2sto} and ~\ref{l2s4sto}, and it appears to be quite robust
in both the $n=1$ and $n=2$ superlattices for a sufficiently large
$J_{\rm AFM}$ (corresponding to narrow bandwidth cases). It should be
noted that the Cx-AFM order does {\it not} exist in bulk form, thus
its stabilization in superlattices would be quite interesting.
The Cx-AFM phase, which breaks the in-plane symmetry between the two directions, will
result in macroscopic physical anisotropies in, e.g., transport properties even when
the underlying substrate is cubic.

The phase diagrams for the case of SrTiO$_3$ as substrate include several other phases
besides the A-AFM and Cx-AFM states. These other phases correspond to combinations
of FM, A-AFM, and G-AFM states, with different patterns, as shown in
Figs.~\ref{l1s2sto} and ~\ref{l2s4sto}.
For configurations  involving a FM component, note that it is quite difficult to obtain
a macroscopic magnetization in these superlattices, since there is no place
for FM order in the phase diagrams for manganites when in bulk form and for hole doping
$x>0.5$.\cite{Tokura:Rpp,Hemberger:Prb02} Thus, if a net magnetic moment could
be stabilized via superlattices for a nominal composition $x$ larger than 0.5,
this would also represent a new state that is not present in bulk form.
Moreover, combinations of FM and G-AFM layers could potentially be
used as spin valves by controlling the local magnetic orders with magnetic fields.

With regards to the LaAlO$_3$ substrate, the C-AFM state is quite robust in
the $n$=$1$ case (Fig.~\ref{l1s2lao}),
which may imply an enhanced N\'{e}el temperature, analogous
to the enhanced A-AFM N\'{e}el temperature observed experimentally
for  SrTiO$_3$ as substrate. However, in the $n$=$2$ case the C-AFM state occupies
a smaller portion of the phase diagram and states such as those involving
$3$ FM layers plus $3$ G-AFM layers may become stable if $V$ is increased further,
similarly as in the case using SrTiO$_3$.
Other combinations such as the $5$ C-AFM layers plus $1$ G-AFM layer
are also possible if  the bandwidth is reduced.
In this case, and also for SrTiO$_3$, the phase diagrams are rich, implying
states that are very close in energy, and it is
possible that external magnetic fields could be used to tune
the ground state properties of these superlattices.


%
\section{Summary}

Summarizing, here the two-orbital double-exchange model was employed to investigate the
phase diagrams of the ($R$MnO$_3$)$_n$/($A$MnO$_3$)$_{2n}$ superlattices.
Two different substrates have
been tested to illustrate two different types of strain effects,
giving rise to distinct ground states and phase diagrams.
The underlying dominant physical mechanism is the uniform
splitting of the $e_{\rm g}$ orbital
levels by the $Q_3$ mode of Jahn-Teller distortion imposed by the substrate strain.
Besides the uniform A-type and C-type antiferromagnetic
phases that are stabilized in the intermediate range
of couplings for the case of the wide-bandwidth
manganite superlattices, our study also reveals several other possible phases, including
many with non-uniform modulations of the spin order. These exotic patterns could be
found in superlattices built using narrow-bandwidth manganites, an area of research
that should be more actively pursued experimentally due to its potential to
unveil new states of manganites.

\begin{acknowledgments}
This work was supported by the 973 Projects of China (2011CB922101), NSFC (11004027), NCET (10-0325), and RFDP (20100092120032). Q.F.Z. and S.Y. were supported by CREST-JST. The work of E.D. for this project
was supported by the U.S. Department of Energy,
Office of Basic Energy Sciences, Materials
Sciences and Engineering Division.
\end{acknowledgments}

\bibliographystyle{apsrev4-1}
\bibliography{../../ref}
\end{document}